\batchmode
\documentstyle[12pt]{article}
\newcommand{\nc}{\newcommand}
\nc{\renc}{\renewcommand}

\nc{\half}{{\textstyle{1\over2}}}
\nc{\etal}{\mbox{\it et al. }}
\nc{\ie}{{\it i.e.}}
\nc{\eg}{{\it e.g.}}

\renc{\thefootnote}{\arabic{footnote}}
\nc{\capt}[1]{{\bf Figure.} {\small\sl #1}}

\nc{\eqs}[2]{\mbox{Eqs.~(\ref{#1},\,\ref{#2})}}
\nc{\eq}[1]{\mbox{Eq.~(\ref{#1})}}

\nc{\figs}[2]{\mbox{Figs.~(\ref{#1},\,\ref{#2})}}
\nc{\fig}[1]{\mbox{Fig~.(\ref{#1})}}

\nc{\tag}[1]{\label{#1} \marginpar{{\footnotesize #1}}}
\nc{\mtag}[1]{\label{#1} \mbox{\marginpar{{\footnotesize #1}}}}
\renc{\baselinestretch}{1.5}
\newlength{\overeqskip}
\newlength{\undereqskip}
\nc{\be}[1]{\begin{equation} \mbox{$\label{#1}$}}
\nc{\bea}[1]{\begin{eqnarray} \mbox{$\label{#1}$}}
\nc{\Section}[2]{\section{#2}\label{#1}}
\nc{\Bibitem}[1]{\bibitem{#1}}
\nc{\Label}[1]{\label{#1}}

\nc{\eea}{\vspace{\undereqskip}\end{eqnarray}}
\nc{\ee}{\vspace{\undereqskip}\end{equation}}
\nc{\bdm}{\begin{displaymath}}
\nc{\edm}{\end{displaymath}}
\nc{\dpsty}{\displaystyle}
\nc{\bc}{\begin{center}}
\nc{\ec}{\end{center}}
\nc{\ba}{\begin{array}}
\nc{\ea}{\end{array}}
\nc{\bab}{\begin{abstract}}
\nc{\eab}{\end{abstract}}
\nc{\btab}{\begin{tabular}}
\nc{\etab}{\end{tabular}}
\nc{\bit}{\begin{itemize}}
\nc{\eit}{\end{itemize}}
\nc{\ben}{\begin{enumerate}}
\nc{\een}{\end{enumerate}}
\nc{\bfig}{\begin{figure}}
\nc{\efig}{\end{figure}}
\nc{\arreq}{&\!=\!&}
\nc{\arrmi}{&\!-\!&}
\nc{\arrpl}{&\!+\!&}
\nc{\arrap}{&\!\!\!\approx\!\!\!&}
\nc{\non}{\nonumber\\*}
\nc{\align}{\!\!\!\!\!\!\!\!&&}

\def\lsim{\; \raise0.3ex\hbox{$<$\kern-0.75em
      \raise-1.1ex\hbox{$\sim$}}\; }
\def\gsim{\; \raise0.3ex\hbox{$>$\kern-0.75em
      \raise-1.1ex\hbox{$\sim$}}\; }
\nc{\DOT}{\hspace{-0.08in}{\bf .}\hspace{0.1in}}
\nc{\Laada}{\hbox {$\sqcap$ \kern -1em $\sqcup$}}
\nc\loota{{\scriptstyle\sqcap\kern-0.55em\hbox{$\scriptstyle\sqcup$}}}
\nc\Loota{{\sqcap\kern-0.65em\hbox{$\sqcup$}}}
\nc\laada{\Loota}
\nc{\qed}{\hskip 3em \hbox{\BOX} \vskip 2ex}

\nc{\real}{{\rm I \! R}}
\nc{\Z}{{\sf Z \!\!\! Z}}
\nc{\complex}{{\rm C\!\!\! {\sf I}\,\,}}
\def\bigid{\leavevmode\hbox{\small1\kern-3.8pt\normalsize1}}
\def\id{\leavevmode\hbox{\small1\kern-3.3pt\normalsize1}}
\nc{\slask}{\!\!\!/}
\nc{\bis}{{\prime\prime}}
\nc{\pa}{\partial}
\nc{\na}{\nabla}
\nc{\ra}{\rangle}
\nc{\la}{\langle}
\nc{\goto}{\rightarrow}
\nc{\swap}{\leftrightarrow}

\nc{\EE}[1]{ \mbox{$\cdot10^{#1}$} }
\nc{\abs}[1]{\left|#1\right|}
\nc{\at}[2]{\left.#1\right|_{#2}}
\nc{\norm}[1]{\|#1\|}
\nc{\abscut}[2]{\Abs{#1}_{\scriptscriptstyle#2}}
\nc{\vek}[1]{{\rm\bf #1}}
\nc{\integral}[2]{\int\limits_{#1}^{#2}}
\nc{\inv}[1]{\frac{1}{#1}}
\nc{\dd}[2]{{{\partial #1}\over{\partial #2}}}
\nc{\ddd}[2]{{{{\partial}^2 #1}\over{\partial {#2}^2}}}
\nc{\dddd}[3]{{{{\partial}^2 #1}\over
        {\partial #2 \partial #3}}}
\nc{\dder}[2]{{{d #1}\over{d #2}}}
\nc{\ddder}[2]{{{d^2 #1}\over{d {#2}^2}}}
\nc{\dddder}[3]{{d^2 #1}\over
        {d #2 d #3}}
\nc{\dx}[1]{d\,^{#1}x}
\nc{\dy}[1]{d\,^{#1}y}
\nc{\dz}[1]{d\,^{#1}z}
\nc{\dl}[1]{\frac{d\,^{#1}l}{(2\pi)^{#1}}}
\nc{\dk}[1]{\frac{d\,^{#1}k}{(2\pi)^{#1}}}
\nc{\dq}[1]{\frac{d\,^{#1}q}{(2\pi)^{#1}}}

\nc{\cc}{\mbox{$c.c.$ }}
\nc{\hc}{\mbox{$h.c.$ }}
\nc{\cf}{cf.\ }
\nc{\erfc}{{\rm erfc}}
\nc{\Tr}{{\rm Tr\,}}
\nc{\tr}{{\rm tr\,}}
\nc{\pol}{{\rm pol}}
\nc{\sign}{{\rm sign}}
\nc{\bfT}{{\bf T }}
\def\eV{{\rm\ eV}}
\def\GeV{{\rm\ GeV}}
\def\MeV{{\rm\ MeV}}

\nc{\cA}{{\cal A}}
\nc{\cB}{{\cal B}}
\nc{\cD}{{\cal D}}
\nc{\cE}{{\cal E}}
\nc{\cG}{{\cal G}}
\nc{\cH}{{\cal H}}
\nc{\cL}{{\cal L}}
\nc{\cO}{{\cal O}}
\nc{\cT}{{\cal T}}
\nc{\cN}{{\cal N}}
\nc{\rvac}[1]{|{\cal O}#1\rangle}
\nc{\lvac}[1]{\langle{\cal O}#1|}
\nc{\rvacb}[1]{|{\cal O}_\beta #1\rangle}
\nc{\lvacb}[1]{\langle{\cal O}_\beta #1 |}
\nc{\bb}{\bar{\beta}}
\nc{\bt}{\tilde{\beta}}
\nc{\ctH}{\tilde{\cal H}}
\nc{\chH}{\hat{\cal H}}
\nc{\1}{\aa}
\nc{\2}{\"{a}}
\nc{\3}{\"{o}}
\nc{\4}{\AA}
\nc{\5}{\"{A}}
\nc{\6}{\"{O}}
\nc{\al}{\alpha}
\nc{\g}{\gamma}
\nc{\Del}{\Delta}
\nc{\e}{\epsilon}
\nc{\eps}{\epsilon}
\nc{\lam}{\lambda}
\nc{\om}{\omega}
\nc{\Om}{\Omega}
\nc{\ve}{\varepsilon}
\nc{\mn}{{\mu\nu}}
\nc{\k}{\kappa}
\nc{\vp}{\varphi}

\nc{\advp}[3]{{\it  Adv.\ in\ Phys.\ }{{\bf #1} {(#2)} {#3}}}
\nc{\annp}[3]{{\it  Ann.\ Phys.\ (N.Y.)\ }{{\bf #1} {(#2)} {#3}}}
\nc{\apl}[3]{{\it  Appl. Phys. Lett. }{{\bf #1} {(#2)} {#3}}}
\nc{\apj}[3]{{\it  Ap.\ J.\ }{{\bf #1} {(#2)} {#3}}}
\nc{\apjl}[3]{{\it  Ap.\ J.\ Lett.\ }{{\bf #1} {(#2)} {#3}}}
\nc{\app}[3]{{\it Astropart.\ Phys.\ }{{\bf #1} {(#2)} {#3}}}
\nc{\cmp}[3]{{\it  Comm.\ Math.\ Phys.\ }{{ \bf #1} {(#2)} {#3}}}
\nc{\cqg}[3]{{\it  Class.\ Quant.\ Grav.\ }{{\bf #1} {(#2)} {#3}}}
\nc{\epl}[3]{{\it  Europhys.\ Lett.\ }{{\bf #1} {(#2)} {#3}}}
\nc{\ijmp}[3]{{\it Int.\ J.\ Mod.\ Phys.\ }{{\bf #1} {(#2)} {#3}}}
\nc{\ijtp}[3]{{\it Int.\ J.\ Theor.\ Phys.\ }{{\bf #1} {(#2)} {#3}}}
\nc{\jmp}[3]{{\it  J.\ Math.\ Phys.\ }{{ \bf #1} {(#2)} {#3}}}
\nc{\jpa}[3]{{\it  J.\ Phys.\ A\ }{{\bf #1} {(#2)} {#3}}}
\nc{\jpc}[3]{{\it  J.\ Phys.\ C\ }{{\bf #1} {(#2)} {#3}}}
\nc{\jap}[3]{{\it J.\ Appl.\ Phys.\ }{{\bf #1} {(#2)} {#3}}}
\nc{\jpsj}[3]{{\it J.\ Phys.\ Soc.\ Japan\ }{{\bf #1} {(#2)} {#3}}}
\nc{\lmp}[3]{{\it Lett.\ Math.\ Phys.\ }{{\bf #1} {(#2)} {#3}}}
\nc{\mpl}[3]{{\it  Mod.\ Phys.\ Lett.\ }{{\bf #1} {(#2)} {#3}}}
\nc{\ncim}[3]{{\it  Nuov.\ Cim.\ }{{\bf #1} {(#2)} {#3}}}
\nc{\np}[3]{{\it  Nucl.\ Phys.\ }{{\bf #1} {(#2)} {#3}}}
\nc{\npps}[3]{{\it  Nucl.\ Phys.\ Proc.\ Suppl.\ }{{\bf #1} {(#2)} {#3}}}
\nc{\pr}[3]{{\it Phys.\ Rev.\ }{{\bf #1} {(#2)} {#3}}}
\nc{\pra}[3]{{\it  Phys.\ Rev.\ A\ }{{\bf #1} {(#2)} {#3}}}
\nc{\prb}[3]{{\it  Phys.\ Rev.\ B\ }{{{\bf #1} {(#2)} {#3}}}}
\nc{\prc}[3]{{\it  Phys.\ Rev.\ C\ }{{\bf #1} {(#2)} {#3}}}
\nc{\prd}[3]{{\it  Phys.\ Rev.\ D\ }{{\bf #1} {(#2)} {#3}}}
\nc{\prl}[3]{{\it Phys.\ Rev.\ Lett.\ }{{\bf #1} {(#2)} {#3}}}
\nc{\pl}[3]{{\it  Phys.\ Lett.\ }{{\bf #1} {(#2)} {#3}}}
\nc{\prep}[3]{{\it Phys.\ Rep.\ }{{\bf #1} {(#2)} {#3}}}
\nc{\prsl}[3]{{\it Proc.\ R.\ Soc.\ London\ }{{\bf #1} {(#2)} {#3}}}
\nc{\ptp}[3]{{\it  Prog.\ Theor.\ Phys.\ }{{\bf #1} {(#2)} {#3}}}
\nc{\ptps}[3]{{\it  Prog\ Theor.\ Phys.\ suppl.\ }{{\bf #1} {(#2)} {#3}}}
\nc{\physa}[3]{{\it  Physica\ A\ }{{\bf #1} {(#2)} {#3}}}
\nc{\physb}[3]{{\it  Physica\ B\ }{{\bf #1} {(#2)} {#3}}}
\nc{\phys}[3]{{\it Physica\ }{{\bf #1} {(#2)} {#3}}}
\nc{\rmp}[3]{{\it  Rev.\ Mod.\ Phys.\ }{{\bf #1} {(#2)} {#3}}}
\nc{\rpp}[3]{{\it Rep.\ Prog.\ Phys.\ }{{\bf #1} {(#2)} {#3}}}
\nc{\sjnp}[3]{{\it Sov.\ J.\ Nucl.\ Phys.\ }{{\bf #1} {(#2)} {#3}}}
\nc{\spjetp}[3]{{\it Sov.\ Phys.\ JETP\ }{{\bf #1} {(#2)} {#3}}}
\nc{\yf}[3]{{\it Yad.\ Fiz.\ }{{\bf #1} {(#2)} {#3}}}
\nc{\zetp}[3]{{\it Zh.\ Eksp.\ Teor.\ Fiz.\  }{{\bf #1}  {(#2)} {#3}}}
\nc{\zp}[3]{{\it Z.\ Phys.\ }{{\bf #1} {(#2)} {#3}}}
\nc{\ibid}[3]{{\sl ibid.\ }{{\bf #1} {#2} {#3}}}
\nc{\rf}[1]{(\ref{#1})}
\nc{\nn}{\nonumber \\*}
\nc{\bfB}{\bf{B}}
\nc{\bfv}{\bf{v}}
\nc{\bfx}{\bf{x}}
\nc{\bfy}{\bf{y}}
\nc{\vx}{\vec{x}}
\nc{\vy}{\vec{y}}
\nc{\oB}{\overline{B}}
\nc{\oI}{\overline{I}}
\nc{\oR}{\overline{R}}
\nc{\rar}{\rightarrow}
\nc{\ti}{\times}
\nc{\slsh}{\hskip-5pt/}
\nc{\sm}{Standard~Model~}
\nc{\MP}{M_{\rm Pl}}
\nc{\tp}{t_{\rm Pl}}
\nc{\ave}{\bar{E}}

\nc{\eff}{{\rm eff}}
\nc{\kk}{\vek{k}}
\nc{\pp}{{\rm p}}
\nc{\ga}{g_{a\gamma}}
\nc{\vv}{\\}
\nc{\eee}{{\bf E}}
\nc{\bbb}{{\bf B}}
\nc{\qcd}{T_{\rm QCD}}
\nc{\G}{\rm \ G}
\def\vec#1{{\bf #1}}

\def\lae{\;^{<}_{\sim} \;} \def\gae{\; ^{>}_{\sim} \;} 

\def\ell{e^{c}LL}

\begin{document}
{\title{\vskip-2truecm{\hfill {{\small \\
        \hfill \\
        }}\vskip 1truecm}
{\LARGE  Thermally Generated Gauge Singlet
 Scalars as Self-Interacting Dark Matter}}
{\author{
{\sc  John McDonald$^{1,2}$}\\
{\sl\small CERN, Theory Division, 1211 Geneva 23, Switzerland}
}
\maketitle
\begin{abstract}
\noindent

           We show that a gauge singlet scalar $S$, with a
 coupling to the Higgs doublet of the
 form $\lambda_{S} S^{\dagger}S H^{\dagger}H$ and with the $S$ mass entirely
 generated by the Higgs expectation value, has a thermally 
generated relic density
$\Omega_{S} \approx 0.3$ if $m_{S} \approx (2.9-10.5)
 (\Omega_{S}/0.3)^{1/5}(h/0.7)^{2/5}\MeV$.  
 Remarkably, this is very similar to the range ($m_{S} =
 (6.6-15.4)\eta^{2/3}$MeV) required in order for the self-interaction 
$(\eta/4)(S^{\dagger}S)^{2}$ to account for self-interacting dark matter 
when $\eta$ is not much smaller than 1. 
The corresponding coupling is $\lambda_{S} 
\approx (2.7 \times 10^{-10} - 3.6 \times
 10^{-9})(\Omega_{S}/0.3)^{2/5}(h/0.7)^{4/5}$, implying that such scalars are
 very weakly coupled to the Standard Model sector. 

\end{abstract}
\vfil
\footnoterule
{\small $^1$mcdonald@sune.amtp.liv.ac.uk}\\
{\small $^2$Mailing address: Theoretical Physics Division,
Dept. of Mathematical Sciences,
University of Liverpool, Liverpool L69 3BX, UK}

\thispagestyle{empty}
\newpage
\setcounter{page}{1}

\section{Introduction}

               It has become apparent that conventional
 collisionless cold dark matter (CCDM)
 may have problems accounting for the
 observed structure of galaxies. N-body simulations 
with CCDM indicate that galaxies should have singular halos 
\cite{sim,moore} 
with large numbers of subhalos \cite{sub,klypin}.
Observationally, the density profile of galaxies in the inner few kiloparsecs 
appears to be much shallower than predicted by numerical simulations
(the central density of dark matter halos being 50 times smaller 
than the CCDM prediction for dwarf galaxies and roughly independent of 
halo mass \cite{moore,firmani1}) whilst the number of dwarf galaxies 
in the Local Group is an order of 
magnitude fewer than predicted \cite{sub,klypin}.
 In addition, the CCDM 
predictions for the Tully-Fisher relation \cite{navarro,tf} and the
 stability of galactic bars in high surface brightness spiral galaxies
\cite{bar} are not in agreement with what is observed, indicating
lower density galaxy cores than predicted by CCDM.
Although there is at present considerable 
uncertainty regarding the interpretation of 
observations and simulations \cite{innerg,bullock},
 it has nevertheless been argued that all
 the discrepencies between observations and simulations 
may be understood as indicating 
that dark matter halos in CCDM simulations are
 more centrally concentrated than observed \cite{ssd}. 

        In order to overcome the possible deficiencies of CCDM halos,
 one suggestion has been that the cold dark matter particles have a 
non-dissipative self-interaction \cite{ss1,ss2}, 
and it has been shown that such
 cold, non-dissipative self-interacting dark matter (SIDM) can be
effective in alleviating the various problems
 of CCDM \cite{ssd}. Scattering of dark matter particles stops 
gravitational accretion at the centre of the halo and so allows a
 smooth core to form. Simulations with
SIDM \cite{ssd,yetal} are able to simultaneously account
 for the observed density
profiles of galactic halo and the number of 
subhalos \cite{ssd}. In the future SIDM may be strongly 
constrained by gravitational lensing observations of the 
shape of cluster halos \cite{lensing,lens2,lensob} 
 and by
 the formation of massive black holes at the 
centres of galaxies, which is enhanced by 
self-interactions of dark matter particles \cite{ost}.

              In order to be able to account for the observed properties of 
dark matter halos, the requirement on the mass $M$ and self-interaction
 scattering cross-section $\sigma$ of the SIDM particles is that 
\cite{ss2}
\be{e1} r_{S} = \frac{\sigma}{M} = (2.05 \times 10^{3} \GeV
 - 2.57 \times 10^{4}) \GeV^{-3}      ~.\ee
The upper bound corresponds to the limit at which galaxy halos in 
massive clusters are destroyed by interacting with hot 
particles in the cluster halo (evaporation) \cite{ss2} 
whilst the lower bound corresponds to the limit where the 
SIDM particle would not interact 
within a typical galactic halo during a Hubble time \cite{ss1,ss2}.

      The canonically simplest dark matter
 particle is arguably a gauge singlet scalar $S$. 
The possibility that gauge singlet scalars, interacting with the
 Standard Model sector via a coupling to the Higgs doublet of the
 form $S^{\dagger}SH^{\dagger}H$,
could naturally constitute dark matter has been
 pointed out by a
 number of authors in the past \cite{zee1,jmcd}
 as well as more recently \cite{burgess}. 
These calculations consider the case of massive
($ > 1 \GeV$) scalars which freeze out of thermal
 equilibrium when non-relativistic \cite{lw}.
 However, the range of $S$
 masses considered is too large to account for 
SIDM with perturbative $S$ couplings. 
  
         It has recently been noted that gauge singlet scalars have a
 natural self-interaction via an $S^{4}$-type coupling and so
 in principle could account for SIDM \cite{bb,bb2,zee2}.
 An estimate of the upper bound on the coupling of $S$ scalars
 to the Higgs doublets for $S$ mass of the order of 10-100 MeV
 (which is of the greatest interest in the
 case of perturbative $S$ self-interactions) 
was derived in \cite{bb2} by requiring that $S$ scalars
 do not come into thermal equilibrium and so overpopulate the Universe. 

      In this letter we 
consider thermal generation of SIDM $S$
 scalars which do not achieve equilibrium. 
 We will show that such non-equlibrium thermal
 generation can naturally account for a dark matter density of $S$ scalars
 with the right properties to account for SIDM. 

        The letter is organized as follows. 
In Section 2 we discuss the perturbative upper limit on 
the $S$ scalar mass. In Section 3 we consider the thermal
 generation of a relic density of $S$ scalars. In Section 4 we consider 
case of zero bare $S$ mass and the resulting consistency of the 
relic density, $S$ mass and Spergel-Steinhardt SIDM cross-section
 for natural values of the $S$ self-coupling. In Section 5 we present our 
conclusions. 

\section{Limit on $m_{S}$ for Perturbative SIDM}

       We first consider the perturbative upper
 limit on the $S$ mass if it is to 
play the role of SIDM. 
We will consider the case of complex 
gauge singlet scalars for consistency
with the cross-sections and discussion
given in \cite{jmcd}, which we will use here. 
We expect that the results for real scalars will be very similar. 
The model is described by 
\be{e2} {\cal L} = \partial_{\mu}S^{\dagger}\partial^{\mu}S
 - m^{2} S^{\dagger}S 
- \lambda_{S}S^{\dagger}SH^{\dagger}H 
 - \frac{\eta}{4} (S^{\dagger}S)^{2} ~.\ee
The total centre-of-mass
 $S$ scattering cross-section is the sum of
 $SS^{\dagger} \rightarrow SS^{\dagger}$ and 
$SS \rightarrow SS$, 
\be{e4} \sigma \equiv 
\sigma_{SS^{\dagger} \rightarrow SS^{\dagger} } + 
\sigma_{SS \rightarrow SS }
= \frac{3\eta^{2}}{128 \pi m_{S}^{2}}    ~.\ee
Therefore 
\be{e5} r_{S} = \frac{\sigma}{m_{S}} = 
\frac{3 \eta^{2}}{128 \pi m_{S}^{3}}     ~,\ee
which implies that 
\be{e6}   m_{S}  = 35.8 \alpha_{\eta}^{1/3} 
\left( \frac{2.05 \times 10^{3} \GeV^{-3}}{r_{S}} \right)^{1/3} \MeV   ~,\ee
where $\alpha_{\eta} = \eta^{2}/4 \pi$. (Similar expressions have been 
obtained in \cite{bb2,zee2}.)
Thus if we require that $\alpha_{\eta} \lae 1$ in 
order to have a perturbative theory then the condition \eq{e1} requires that
 $m_{S}  = \alpha_{\eta}^{1/3}(15.4-35.8) \MeV
\lae 30 \MeV$. (We refer to this as the Spergel-Steinhardt
 mass range.) 
We note that this puts a severe bound on the coupling $\lambda_{S}$, since 
the $S$ scalar gains a mass from the Higgs expectation value, 
\be{e7} m_{S}^{2} = m^{2} + \frac{\lambda_{S} v^{2}}{2}     ~,\ee
where $v = 250 \GeV$. 
Thus the requirement that $m_{S} \lae 30 \MeV$
 imposes an upper bound on $\lambda_{S}$
\be{e8}   \lambda_{S}  <  
  2.9 \times 10^{-8} \left(\frac{m_{S}}{30 \MeV}\right)^{2}    ~.\ee
The Spergel-Steinhardt mass range assumes a perturbative $S$ 
self-coupling. A non-perturbative self-coupling may be
 possible, but it would be difficult to calculate 
the properties of such a model, 
so we must restrict ourselves to the perturbative case. In addition, 
the only known scalar self-coupling, that of the Standard Model Higgs doublet, 
$\lambda_{H} (H^{\dagger}H)^{2}$, is given by $\lambda_{H} = 
m_{h}^{2}/4 v^{2} = 0.053 (m_{h}/115 \GeV)^{2}$. (The experimental lower bound 
on the Higgs mass is $m_{h} > 113 \GeV$ 
\cite{hm}, whilst  
 an upper bound for the pure Standard Model is  
obtained from radiative corrections to
 electroweak observables, $m_{h} < 165 \GeV$ \cite{hm2}. 
The upper bound in extensions
 of the Standard Model can be 1 TeV or larger \cite{frs}.) 
This is typically perturbative but
 not very much smaller than 1, suggesting that a natural 
value for the $S$ scalar
 self-couplings is around 0.1.  

\section{Thermal Generation of $S$ Scalars}

       There are two processes which can 
produce a density of $S$ scalars: 
2 $\leftrightarrow$ 2 annihilation processes and decay
 of a thermal equilibrium density of Higgs scalars to $SS^{\dagger}$ pairs, 
$h^{o} \rightarrow SS^{\dagger}$. We first consider 2 $\leftrightarrow$ 2 
annihilations. The relic density from scattering processes
 in a radiation dominated Universe is found by solving
 the Boltzmann equation \cite{lw,jmcd}, 
\be{e12} \frac{df}{dT} = 
\frac{<\sigma_{ann}v_{rel}>}{K}(f^{2} - f_{o}^{2}) \;\;\; ; \;\;  
 K = \left[ \frac{4 \pi^{3} g(T)}{45 M_{Pl}^{2}} \right]^{1/2}   ~\ee
where $f = n_{S}/T^{3}$, $f_{o} = n_{o}/T^{3}$ and
 $g(T) = g_{B} + 7g_{F}/8$, where $g_{B}$
 and $g_{F}$ denote the number of relativistic
 bosonic and fermionic degrees of freedom respectively. 
$n_{S}$ is the number density of $S$ scalars
 and $n_{o}$ is the thermal equilibrium $S$ number
 density; for relativistic $S$ scalars  
\be{e11a} n_{o} = \left(\frac{1.2}{\pi^{2}}\right)T^{3}   ~.\ee
We consider the case where the $S$ scalar
 density is very small compared with
 the equilibrium density and solve the Boltzmann
 equation with $f=0$ on the
 right-hand side, 
\be{e17} \frac{df}{dT} = 
- \frac{<\sigma_{ann}v_{rel}>}{K} f_{o}^{2}   ~.\ee
We take the Universe to be initially at a high
 temperature, $T \gg m_{W}$, and calculate the
 resulting relic density of $S$ scalars as the Universe cools.
 The annihilation cross-sections of relativistic 
$SS^{\dagger}$ pairs to $t$ quarks, $W$ and $Z$ bosons and the
 $h^{o}$ Higgs scalars
 (lighter quarks and leptons do not contribute
 significantly due to their very small 
Yukawa couplings) are estimated by using the centre-of-mass
 annihilation cross-sections 
calculated for $S$ scalars with typical energy $E_{T} \approx T$.
 We will see that the core results
 of the paper are not very sensitive to uncertainties
 in the calculation of the annihilation cross-section and
 thermal relic $S$ density. The relativistic annihilation cross-sections
 $\sigma_{i}$ may be obtained from the non-relativistic 
$<\sigma_{ann}v_{rel}>$ given in \cite{jmcd} via
 the relation $\sigma_{i} = 
(1/2)<\sigma_{ann}v_{rel}>(m_{S} \rightarrow E_{T})$ (where $i \equiv 
{t,W,Z,h^{o}}$ denotes the Standard Model particle in question), which may
 be confirmed by directly calculating the cross-sections. For $T < m_{i}$
the contribution of $\sigma_{i}$ to the total cross-section is
 zero, which models Boltzmann suppression.
Then $<\sigma_{i}v_{rel}> = 2\sigma_{i}$, where we take $v_{rel} = 2$ 
for relativistic annihilations \cite{gg}. 
For relativistic
 $S$ scalars, $f_{o} = 1.2/\pi^{2}$ is a constant 
so \eq{e17} can be integrated as
\be{e22} f_{i} = - 2f_{o}^{2} 
\int_{E_{T_{o}}}^{E_{T_{f}}} dE_{T}
 \; \frac{\sigma_{i}}{K}    ~,\ee
where $E_{T} = T$, $E_{T_{f}} = m_{i}$ 
and the initial thermal energy $E_{T_{o}} \rightarrow \infty$. 
We will take $K \propto g(T)^{1/2}$ to be 
constant with $g(T) = g(T_{i})$, where 
$T_{i} = m_{i}$, since most of the integral comes 
from $E_{T}$ close to $m_{i}$.
The total contribution to $f$ is then
\be{e44}   f_{T} = \sum f_{i} = 
1.3 \times 10^{12} \lambda_{S}^{2} \left(1 + 0.27 
\left(\frac{115 \GeV}{m_{h}}\right) + 
 0.20 \left(\frac{m_{h}}{115 \GeV}\right)^{2} \right) 
~.\ee
In this we have taken $g(T_{i}) = 106.75$, corresponding
 to the Standard Model degress of
 freedom and $\lambda_{t} = 0.7$ (corresponding to 
$m_{t} =175 \GeV$). 
In addition, in order to obtain an analytical result we have expanded
 the Higgs propagators in $\sigma_{i}$ ($i= W,Z,t$) assuming that 
$4E_{T}^{2}$ is large compared with
 $m_{h}^{2}$, which is generally satisfied if 
$m_{h}^{2}$ is small compared with $4m_{W}^{2}$. (We refer to this as 
the small Higgs mass limit.)  

     The $S$ number density from the decay of thermal
 equilibrium $h^{o}$ scalars 
at temperatures less than the electroweak phase
 transition (where $T_{EW} \gae 1.5 m_{h}$ \cite{jansen}) is given by 
\be{x1} \frac{dn_{S}}{dt} + 3 H n_{S}
 = <\Gamma_{h^{o}}> n_{h^{o}\; eq}   ~,\ee
where $H$ is the expansion rate and 
the thermal equilibrium density of
 $h^{o}$, $n_{h^{o}\;eq}$, is given by 
\be{x2} n_{h^{o} \; eq} = \frac{1}{2 \pi^{2}}
 \int_{m_{h}}^{\infty} \frac{E \left(E^{2}
 - m_{h}^{2}\right)^{1/2}}{\left(e^{E/T} -1\right)} dE    ~\ee
and the decay rate for
 $h^{o}$ scalars with energy $E$ is 
\be{x4}   \Gamma_{h^{o}} =
 \frac{\lambda_{S}^{2}v^{2}}{16 \pi E}    ~.\ee
Thus the thermal average of the decay rate is 
\be{x6}  < \Gamma_{h^{o}}> = 
\frac{1}{n_{h^{o}\;eq}}
 \frac{\lambda_{S}^{2}v^{2}T^{2}
e^{-m_{h}/T}\eta(m_{h}/T)}{32 \pi^{3}} \;\;\; ; \;\; 
\eta(a) = \int_{0}^{\infty} \frac{t^{1/2}
 \left(t + 2 a \right)^{1/2}}{\left(e^{t} -e^{-a}\right)} dt 
~.\ee
Therefore in terms of $f$, the $S$ density from $h^{o}$ decays is given by
\be{x8} \frac{d f}{d T} = 
   -\frac{<\Gamma_{h^{o}}>f_{h^{o}\;eq}}{K T^{3}} 
\equiv -\frac{\eta(m_{h}/T)}{KT^{4}}
\frac{\lambda_{S}^{2}v^{2}e^{-m_{h}/T}}{32 \pi^{3}}  ~.\ee
$\eta(a)$ is a slowly varying function of $a$, 
with $\eta(0) = 1.64$, $\eta(1) = 1.87$
and $\eta(5) = 3.00$. Since most of the contribution to $f$ comes
 from $m_{h}/T \sim 1$, we take $\eta(m_{h}/T)$ to be
 equal to $\eta(1)$, in order to obtain an analytical expression. Therefore the
 density of $S$ scalars from $h^{o}$ decay, $f_{dec}$, is given by, 
\be{x9} f_{dec} = \frac{\lambda_{S}^{2}
v^{2}\eta(1)}{16 \pi^{3}  K m_{h}^{3}} \approx 
= 1.08 \times 10^{14}
\lambda_{S}^{2}
 \left(\frac{115 \GeV}{m_{h}}\right)^{3}  
 ~.\ee
In this we have assumed that $v$ is given by
 its $T=0$ value, $v=250 \GeV$. Since 
most of the contribution to $f_{dec}$ comes
 from $T \lae m_{h} < T_{EW}$, this should be a reasonable approximation. 
We see that the $S$ density from $h^{o}$ decays
 is generally much larger than that from 
$2 \leftrightarrow 2$ annihilation processes in the small Higgs mass limit. 
For larger values of the Higgs mass, it 
is possible that s-channel pole annihilations \cite{gg}
may result in $2 \leftrightarrow 2$ 
processes dominating the $h^{o}$ decays 
\cite{jmcd2}, in which case $f_{dec}$ is a lower bound on 
the number of $S$ scalars produced thermally. 

     The resulting density of $S$ plus $S^{\dagger}$ scalars is then 
the sum of scattering and decay contributions 
\be{e40} \Omega_{S} = 
\frac{2 m_{S}}{\rho_{c}} g(T_{\gamma}) T_{\gamma}^{3} 
\sum \frac{f_{i}}{g\left(T_{i}\right)}    ~\ee
where $\rho_{c} = 7.5 \times 10^{-47} h^{2} \GeV^{4}$
 is the critical density, $T_{\gamma} = 2.4 \times 10^{-4} \eV$ is the 
present photon temperature, 
 $T_{i} \approx m_{i}$ and $g(T_{\gamma}) = 2$. 
In this we have used the 
fact that the $S$ number density to entropy is conserved
 once the scattering and decay 
processes are Boltzmann suppressed, such that 
$g(T) n_{S}/T^{3}$ is constant. 
Therefore with $f 
\approx f_{dec}$ and $g(T_{i}) = 106.75$ 
for $T_{i} \approx m_{h}$, the thermal relic 
$S$ density $\Omega_{S}$ is related to $\lambda_{S}$ by 
\be{e46} \lambda_{S}  = 2.0 \times 10^{-10} \frac{h}{\eta^{1/3}} 
\left(\frac{\Omega_{S}}{0.3}\right)^{1/2} 
\left(\frac{10 \eta^{2/3} \MeV}{m_{S}}\right)^{1/2}
 \left(\frac{m_{h}}{115 \GeV}\right)^{3/2}    ~.\ee
Thus for Higgs masses
 in the range 115 GeV to 1 TeV
and with expansion rate $h \approx 0.7$, 
the upper bound on $\lambda_{S}$ from 
requiring that $\Omega_{S} \lae 0.3$ is in the range 
$(1.4 \times 10^{-10}- 3.6 \times 10^{-9}) \eta^{-1/3}$. 
This is in broad agreement with the upper bound estimated in \cite{bb2}, 
based on the 
weaker condition that the $S$ scalars do not come into thermal equilibrium. 
More importantly, we see that it is possible to generate a thermal relic 
density with $\Omega_{S} \approx 0.3$ and $m_{S} \approx 10 \MeV$ 
(typical of SIDM scalars) purely within the
 minimal gauge singlet scalar extension of the Standard Model.

\section{Naturally Consistent Thermal Relic 
SIDM for Zero Bare Mass}

          The value of $\lambda_{S}$ from requiring that 
$\Omega_{S} \approx 0.3$ is satisfied is not very much smaller than the 
 upper limit \eq{e8} from the requirement that the Higgs
 expectation value contribution to the $S$ mass
 is compatible with
perturbative SIDM $S$ scalars. 
 This suggests that it is quite likely that {\it all}
 the $S$ mass might come from 
its interaction with the Higgs scalar when its
relic density is sufficient to account for dark matter.
 If we assume that all the $S$ mass is due to the Higgs
 expectation value, then 
we find that the $S$ mass is {\it fixed} by the thermal relic density
\be{e47} m_{S} = 2.9 \left(\frac{\Omega_{S}}{0.3}\right)^{1/5} 
\left(\frac{h}{0.7}\right)^{2/5} 
\left(\frac{m_{h}} {115 \GeV}\right)^{3/5} \MeV      ~.\ee 
We refer to this as the thermal relic $S$ mass.
 Comparing with the Spergel-Steinhardt range for SIDM,
\be{e48} m_{S} = (6.6-15.4) \eta^{2/3} \MeV    ~,\ee
we see that the thermal relic mass for $S$ scalars
 is within the range
 required to account for SIDM when the
 self-coupling constant $\eta$ is equal to
 about 0.1, a natural value which is consistent with the Higgs doublet
 self-coupling in the Standard Model. The thermal relic
 mass is not strongly dependent upon cosmological
 parameters, nor is it strongly dependent
 upon the Higgs mass. 
 In particular, it is relatively insensitive to uncertainties in the 
calculation of $f$, since a change in $f$ by a factor 
$\delta$ produces a change in $\Omega_{S}$ by
 the same factor, and so a change in the
thermal relic mass by $\delta^{1/5}$.
 The coupling corresponding to the thermal relic mass is 
\be{e49} \lambda_{S} = 2.7 \times 10^{-10} 
 \left(\frac{\Omega_{S}}{0.3}\right)^{2/5} 
\left(\frac{h}{0.7}\right)^{4/5} \left(\frac{m_{h}} {115 \GeV}\right)^{6/5} 
~.\ee
This suggests a scenario for dark matter in which
 stable gauge singlet scalars couple very weakly to the
 Standard Model sector but self-couple with a relatively
strong coupling of about 0.1, of the order expected from the 
example of the Standard Model Higgs self-coupling.

\section{Conclusions}

         We have considered the thermal generation of a relic 
density of self-interacting dark matter (SIDM) gauge singlet
 scalars. The dominant process for small Higgs mass 
is the decay of thermal 
equlibrium Higgs scalars to gauge singlet scalar pairs.
For SIDM scalars with
 perturbative self-interactions, the mass 
must be no greater than around $30 \MeV$.
 For such light scalars, the requirement of an
 acceptable relic density of $S$ scalars requires
 that the $S$ coupling to the Standard Model 
Higgs satisfies $\lambda_{S} \lae
 10^{-(9-10)}$. This limit comes from the requirement that 
$S$ scalars are not thermally overproduced, which is
 a stronger condition than requring that they do not 
come into thermal equilibrium. 
 
          In the case where $S$ scalars account for dark matter 
and where the $S$ mass is entirely due
 to the Higgs expectation value,
 we find that the $S$ mass is fixed by the thermal
 relic dark matter density to be between 
about 2.9 MeV and 10.5 MeV for Higgs masses ranging
 from 115 GeV to 1 TeV. (The upper limit on the $S$ mass 
may be smaller if $2 \leftrightarrow 2$ 
annihilations dominate $h^{o}$ decays for large Higgs mass.) 
This is very similar to  
the range of masses ( $(6.6-15.4)\eta^{2/3}$MeV) required by
 self-interacting dark matter with self-coupling $\eta$
 of the order of the natural value 
(based on comparison with the Standard Model Higgs
 doublet self-coupling) of around 0.1. 
 This result is not strongly sensitive to uncertainties either
 in the cosmological parameters or in the calculation of the
 thermal relic $S$ density. 
We find this coincidence
 remarkable and a possible hint that light gauge singlet scalars
 with very weak coupling to the Standard Model sector may
 play an important role in cosmology and particle physics.

        The Author would like to thank the CERN theory division for its 
hospitality and PPARC for a Travel Fund award.

\end{document}